\documentclass[conference]{IEEEtran}
\IEEEoverridecommandlockouts
\usepackage{graphics} 
\usepackage{fancybox}
\usepackage{multirow}
\usepackage{flushend}
\usepackage{booktabs}
\usepackage{tabularx}
\usepackage{comment}
\usepackage{array}
\usepackage[flushleft]{threeparttable}
\usepackage{mdframed}
\graphicspath{{}{images/}{dia/}}
\DeclareGraphicsExtensions{.pdf,.png}

\usepackage{graphicx}
\usepackage{subfig}

\usepackage{listings}
\usepackage{courier}
\usepackage{hyperref}
\usepackage[normalem]{ulem}
\usepackage{color}
\usepackage{colortbl}
\usepackage[table]{xcolor}
\usepackage{hyphenat}

\usepackage{cite}
\usepackage{amsmath,amssymb,amsfonts}
\usepackage{algorithmic}
\usepackage{graphicx}
\usepackage{textcomp}
\usepackage{xcolor}
\usepackage{adjustbox}
\usepackage{multirow}
\usepackage{booktabs}
\def\BibTeX{{\rm B\kern-.05em{\sc i\kern-.025em b}\kern-.08em
    T\kern-.1667em\lower.7ex\hbox{E}\kern-.125emX}}

\usepackage{tikz}
\usepackage{pgfplots}
\usepackage{pgf-pie}
\usepackage[most]{tcolorbox}
\usepgfplotslibrary{colorbrewer}
\usepackage{babel}
\usepackage{wrapfig}
\pgfplotsset{compat = 1.15, cycle list/Set1-8} 
\pgfplotsset{
    /pgfplots/custom legend/.style={
    legend image code/.code={
        \draw[only marks,mark=square*] plot coordinates {(0.3cm,0cm)};
        },
    },
}
\usetikzlibrary{pgfplots.statistics, pgfplots.colorbrewer} 

\usepackage{xcolor,colortbl}
\usepackage{pgfplotstable}
\usepackage{url}
\definecolor{LightBlue}{RGB}{255, 226, 215}

\begin{document}

\title{How Do Community Smells Influence Self-Admitted Technical Debt in Machine Learning Projects?\\}

\author{Shamse Tasnim Cynthia \hspace{4mm} Nuri Almarimi \hspace{4mm}  Banani Roy\\
\normalsize Department of Computer Science, University of Saskatchewan, Canada\\
\normalsize \{shamse.cynthia, nuri.almarimi,  banani.roy\}@usask.ca
}
\IEEEoverridecommandlockouts
\IEEEpubid{\makebox[\columnwidth]{979-8-3315-9147-2/25/\$31.00~\copyright2025 Crown IEEE \hfill}
\hspace{\columnsep}\makebox[\columnwidth]{ }}

\maketitle
\IEEEpubidadjcol
\begin{abstract}
Community smells reflect poor organizational practices that often lead to socio-technical issues and the accumulation of Self-Admitted Technical Debt (SATD). While prior studies have explored these problems in general software systems, their interplay in machine learning (ML)-based projects remains largely underexamined. In this study, we investigated the prevalence of community smells and their relationship with SATD in open-source ML projects, analyzing data at the release level. 
First, we examined the prevalence of ten community smell types across the releases of 155 ML-based systems and found that community smells are widespread, exhibiting distinct distribution patterns across small, medium, and large projects.
Second, we detected SATD at the release level and applied statistical analysis to examine its correlation with community smells. Our results showed that certain smells, such as Radio Silence and Organizational Silos, are strongly correlated with higher SATD occurrences. 
Third, we considered the six identified types of SATD to determine which community smells are most associated with each debt category. Our analysis revealed authority- and communication-related smells often co-occur with persistent code and design debt. 
Finally, we analyzed how the community smells and SATD evolve over the releases, uncovering project size-dependent trends and shared trajectories. 
Our findings emphasize the importance of early detection and mitigation of socio-technical issues to maintain the long-term quality and sustainability of ML-based systems.
\end{abstract}


\section{Introduction}

Community smells arise from poor organizational and social practices that contribute to the buildup of social debt\cite{tamburri2015social, 9402214, Almarimi2020LearningTD}. Social debt leads to communication breakdowns, weak collaboration, and coordination issues \cite{8651329, 9402214}. Such socio-organizational deficiencies are known as ``community smells". If left unaddressed, these sub-optimal patterns can increase project complexity and cost, ultimately leading to the accumulation of technical debt and impacting software quality~\cite{5521754}.
%
These challenges are equally relevant in machine learning (ML) projects  where the human and collaborative aspects of development remain underexplored. As a result, teams often face difficulties in coordinating work, defining interfaces, and managing system evolution~\cite{10.1145/3510003.3510209}.
Nahar et al. \cite{10.1145/3510003.3510209} identified factors such as cultural differences, power struggles, and uneven expertise as key contributors to collaboration breakdowns in ML-based systems. These socio-technical issues can, in turn, lead developers to take shortcuts, apply quick fixes, or defer proper solutions, which they often document explicitly in the form of Self-Admitted Technical Debt (SATD)\cite{zampetti2021self, Potdar2014}. 
Disputes, discomfort, or conflicts among developers due to socio-technical factors can significantly contribute to the emergence of SATDs during project development. The more technical debt persists, the more expensive it becomes, demanding substantial time, effort, and knowledge from developers to resolve. 
Therefore, to better understand the interplay between organizational dynamics and SATD in ML projects, this study aims to: (1) examine the prevalence of community smells in ML-based systems, (2) explore their co-occurrence with SATD, (3) identify which SATD types are most associated with specific community smells, and (4) analyze how both evolve across releases in projects of different sizes.

\begin{table*}[]
    \centering
    \caption{Details of Community Smells}
    \normalsize
    \resizebox{0.7\textwidth}{!}{
    \begin{tabular}{|p{5cm}|p{16cm}|}
    \toprule
         \textbf{Name} & \textbf{Definition}  \\
         \midrule
         \textbf{Organizational Silo Effect (OSE)} & This smell refers to distinct subgroups within a community and poor communication among developers, which leads to wasted time, duplicated code, and added project costs.\cite{10.1109/52.819967, tamburri2015social, 8651329}.  \\  \midrule
         \textbf{Black-cloud Effect (BCE) } & This smell reflects information overload due to poor communication, limited knowledge-sharing opportunities, and a lack of experienced members to bridge gaps in expertise. \cite{tamburri2015social, 8651329}. \\ \midrule
         \textbf{Prima-donnas Effect (PDE) } & This smell originates when a group of individuals is resistant toward change proposed by other team members due to poorly organized collaboration. \cite{10.1109/52.819967, tamburri2015social, 8651329}. \\ \midrule
         \textbf{Sharing Valley (SV)} & This smell stems from a lack of effective information exchange, such as face-to-face meetings, that leads to the spread of outdated or unreliable knowledge. \cite{tamburri2015social}. \\
         \midrule
         \textbf{Organizational Skirmish (OS) } & This smell arises from mismatched expertise and communication gaps among team members, often leading to lower productivity and affecting the project's timeline and budget. \cite{tamburri2015social}. \\ \midrule
         \textbf{Solution Defiance (SD) } & This smell arises from cultural and experiential differences, leading to opposing subgroups with conflicting views on technical and socio-technical decisions. The presence of this smell often results in project delays and uncooperative behavior. \cite{tamburri2015social}. \\ \midrule
         \textbf{Radio Silence (RS) }  & This smell originates from excessive formality and poor organizational structure, slowing down changes and wasting time. It often causes significant delays in decision-making due to the rigidity of formal procedures \cite{tamburri2015social}. \\ \midrule
         
         \textbf{Truck Factor Smell (TFS) } & The smell occurs when critical knowledge is concentrated in a few developers, leading to significant knowledge loss if they leave the project. \cite{avelino2016novel}.\\ \midrule
         
         \textbf{Unhealthy Interaction (UI) } & This smell arises when slow, shallow or low-quality discussions, with minimal developer engagement and long delays, take place between message exchanges. \cite{9397528}. \\ \midrule
         
        \textbf{Toxic Communication (TC) } & This smell is present when developer communications involve harmful dialogues and express negative emotions, including dissatisfaction, anger, or discordant viewpoints on different issues under discussion. Such interaction can cause frustration and may lead developers to disengage from the project. \cite{9397528, 10.1007/s10664-017-9526-0}.\\
        \bottomrule
    \end{tabular}
    }
    
    \label{tab:community-smell-details}
    \vspace{-1em}
\end{table*}
Prior studies have explored the relationship between community smells and various technical and social aspects of software development \cite{palomba2018beyond, palomba2018community, tamburri2019splicing, mumtaz2022analyzing, 10172624, zampetti2021self}. For instance, Palomba et al. \cite{palomba2018beyond} linked code and community smells, emphasizing the need for active communication among developers to enhance maintainability and reduce technical debt. Tamburri et al. \cite{tamburri2019splicing} found that community smells often align with sub-optimal architectural structures, while Mumtaz et al. \cite{mumtaz2022analyzing} explored their relationship with design smells. On the other hand, in the context of ML systems,  Mailach et al.\cite{10172624} identified leadership gaps, organizational silos, and poor communication as key socio-organizational challenges,
and Zampetti et al.\cite{zampetti2021self} showed that organizational and project cultures significantly influence developers in introducing SATD \cite{Potdar2014}. Although Annunziata et al. \cite{annunziata2024uncovering} have explored community smells in ML projects, they typically focused on existing literature and extended it through developer interviews and supporting statistical models. Additionally, no studies have explored the prevalence of community smells in the ML systems at the release level. While previous work highlights socio-technical challenges in ML development, the interplay between community smells and SATD over project evolution remains underexplored. 

In this study, we addressed these gaps through an empirical investigation of 155 open-source ML projects, analyzing them at the release level. We aimed to detect \textit{ten} types of community smells and their prevalence in the ML projects. Additionally, we conducted a correlation analysis to explore the relationship between community smells and SATD, and further examined which types of community smells are most associated with specific \textit{six} SATD categories. Furthermore, we analyzed how both community smells and SATD evolve across project releases in ML systems of varying sizes. 
In particular, this study answered \textit{four} research questions and made \textit{four} key contributions, as outlined below:

\textbf{RQ1: To what extent are community smells prevalent in ML-based systems?} \\
We examine the prevalence of various community smells in ML projects to understand how socio-technical issues manifest in these systems. This analysis is essential for identifying early collaboration bottlenecks, guiding organizational improvements, and fostering healthier ML development practices.

\textbf{RQ2: To what extent do community smells co-occur with SATD in ML-based systems?}\\
We investigate the relationship between community smells and the presence of SATD in ML projects. Understanding this co-occurrence helps uncover how socio-technical issues contribute to technical debt accumulation and informs targeted strategies for mitigating both.

\textbf{RQ3: What types of SATD are most associated with specific community smells?}\\
We analyze the associations between individual community smells and distinct types of SATD to identify which socio-technical issues most strongly contribute to specific forms of technical debt. This insight supports more focused interventions to address the root causes of recurring debt patterns in ML-based systems.

\textbf{RQ4: How do community smells and SATD evolve over releases in ML projects of different sizes?} 
We examine the temporal trends of community smells and SATD across small, medium, and large ML projects to understand how these issues progress throughout a project’s lifecycle. This analysis highlights size-specific patterns and informs proactive debt management and collaboration strategies over time.

\textbf{Replication Package} is available in our online appendix \cite{replication_package}.
\vspace{-1em}
\section{Background}
\textbf{Community Smells:} ``Community smell" is a term within an advanced research field that delineates socio-technical characteristics (e.g., high formality) and patterns (e.g., repeated condescending behaviour, rage quitting) leading to the emergence of social debt \cite{Tamburri2016TheAR}.  
Previous studies have identified a range of community smells that reflect socio-technical issues within development teams. In our research, we refer to the ten types of community smells detected by the csDetector tool \cite{10.1145/3468264.3473121}. Table~\ref{tab:community-smell-details} summarizes the definition of ten different types of community smells studied in our research.

\begin{figure*}
    \centering
    \includegraphics[width=0.8\textwidth]{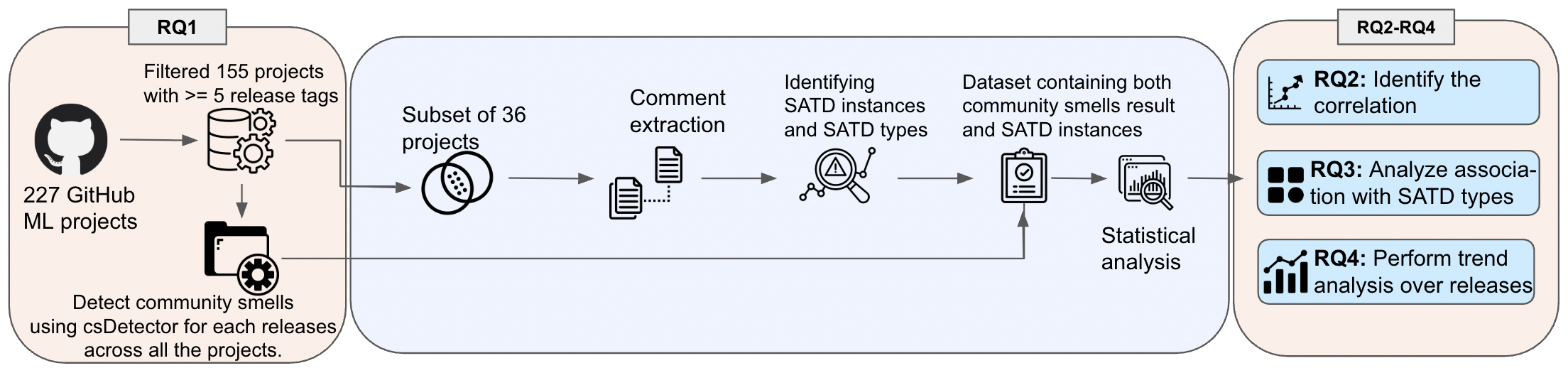}
    \caption{Schematic diagram of our methodology}
    \label{fig:methodology}
    \vspace{-1em}
\end{figure*}

\textbf{Self-admitted Technical Debt:}
Ward Cunningham initially introduced the concept of technical debt in 1993 as a metaphor, highlighting the inevitable costs, such as maintenance and evolution expenses, that developers incur when dealing with code that is not quite optimal \cite{10.1145/157709.157715}. Technical debt often arises due to tight deadlines, high-quality demands, or the need for rapid delivery, which compel developers to make short-term trade-offs \cite{10.1145/1882362.1882373, 6336722, 6280547,7332619}. When developers are aware that their short-term implementations are suboptimal, they may signify this by adding source code comments, a phenomenon recognized as SATD \cite{Potdar2014}. Prior research by Wehaibi et al. \cite{7476641} indicates that the presence of SATD in certain projects undergoes more bug-fixing changes, making these debt-related modifications more challenging.

\section{Related Work}
\textbf{Community smells in software repositories:} The influence of community smells on various aspects of software development has been extensively studied \cite{8546762,9825896,10.1016/j.infsof.2022.107078,almarimi2023improving,Almarimi2020LearningTD}. Palomba et al.~\cite{8546762} found that developers often perceive a connection between code and community smells. Sarmento et al.~\cite{9825896} examined the role of gender diversity and refactoring in mitigating these smells. Caballero-Espinosa et al.~\cite{10.1016/j.infsof.2022.107078} framed community smells as social debt and reviewed their implications. Almarimi et al.~\cite{Almarimi2020LearningTD,almarimi2023improving} proposed machine learning-based detection models that leverage socio-organizational patterns and later extended this to multi-label detection of eight smell types \cite{10.1145/3372787.3390439}. Similarly, Voria et al.~\cite{9978263} introduced CADOCS, a conversational agent built on community smell detection tools.
However, in the context of ML-based systems, community smells have not been studied comprehensively. A recent study by Annunziata et al. \cite{annunziata2024uncovering} tried to identify the types, causes, effects and potential mitigation strategies of community smells in ML-enabled systems. However, their work is built on existing research by extending it through interviews and reinforcing it with statistical models. 
Conversely, Nahar et al. \cite{10.1145/3510003.3510209} highlighted that human collaboration factors are often overlooked in ML projects, despite posing significant challenges. Through interviews with 45 developers, they examined issues related to team structure, cross-role interaction, and conflict across platforms. 
Moreover, Mailach et al. \cite{10172624} identified 17 non-technical anti-patterns tied to poor communication and organizational gaps. Other studies confirm that interdisciplinary ML teams often face collaboration barriers due to cultural, methodological, and expectation mismatches \cite{brandstadter2016interdisciplinary, brown1995factors, passi2020making}.
Given the clear impact of poor organizational and social practices on ML-based development, our study investigates the prevalence of community smells in ML systems, focusing on the specific smell types that have not been explored in this context before.

\textbf{SATD in different sources:} 
Sierra et al. \cite{SIERRA201970} have highlighted that developers acknowledge the presence of SATD from various sources. 
Zampetti et al. \cite{zampetti2021self} surveyed 101 developers and found that source code comments are the most common method for documenting SATD in both industrial and open-source projects. 
Additionally, several empirical studies have investigated SATD. Potdar and Shihab \cite{Potdar2014} analyzed source code comments from four large open-source projects and found that SATD prevalence ranged from 2.4\% to 31\% across files. Their study also revealed that more experienced developers were responsible for introducing most of the SATD. 
Moreover, a wide range of research has been done to identify SATD in the software repositories over the years \cite{10.1145/3524842.3528469, maipradit2020wait, de2020identifying}. 
Researchers have also explored SATD in ML-based systems. O’Brien et al. \cite{10.1145/3540250.3549088} conducted a large-scale study on technical debt in ML software, analyzing 68,820 SATD instances across 2,641 curated GitHub repositories. Their findings revealed five new and five split SATD types specific to ML projects. 
Not only that, Bhatia et al. \cite{bhatia2023empirical} further found that ML projects have a median SATD percentage that is twice as high as that of non-ML projects. 
However, to the best of our knowledge, no prior study has thoroughly investigated the reasons behind the high prevalence of SATD in ML-based projects. Given that community smells often signal underlying technical and organizational issues, our study aims to explore whether a correlation exists between community smells and SATD instances and their evolution over the time. 

\section{Methodology}
This section describes our study research approach to answer the research questions. Fig.~\ref{fig:methodology} shows the schematic diagram of our study methodology.
\subsection{Dataset Description}
The first step of our study involved selecting appropriate ML projects. To this end, we built upon the work of Latendresse et al. \cite{latendresse2024exploratory} for two primary reasons. First, their contribution provides a curated dataset of 227 open-source ML projects, offering a robust foundation for our analysis. 
These projects were carefully selected to represent active, high-quality, and diverse repositories. Second, their selection criteria ensured that the chosen projects are mature and stable in their development lifecycle characteristics. These criteria are crucial for our investigation, as mature projects are more likely to have experienced and responded to various challenges, thereby offering a richer and more meaningful dataset to investigate the prevalence of community smells and their relationship with SATD. Given our focus on community-related smells, this dataset aligns well with the objectives of our study. 

To investigate the relationship between community smells and SATD, we needed to observe both within the same period of project development, i.e., they happen at the same time or close to each other. For this reason, we chose to analyze the presence of both patterns at the release level \cite{8546762}.\begin{wraptable}[6]{r}{0.30\textwidth}
    \captionsetup{font=small}
    \centering
    \vspace{-6pt}
    \resizebox{.30\textwidth}{!}{
    \begin{tabular}{l|c|c|c}
        \hline
        Type     &\begin{tabular}{@{}c@{}}Community\\Dimension\end{tabular}& \begin{tabular}{@{}c@{}}\# of\\Projects\end{tabular} & \begin{tabular}{@{}c@{}}\# of\\releases\end{tabular} \\
        \hline
        Small   &   $<$ 50      &   81      &   3081\\
        Medium  &  50 $-$ 150   &   44      &   1867\\
        Large   &  $>$ 150      &   30      &   1521\\ 
        \hline
    \end{tabular}
    }
    \caption{Breakdown of the projects}
    \label{tab:project_breakdown}
\end{wraptable} We filtered the original 227 projects to keep only those with at least five tagged releases to ensure a richer dataset with enough information for meaningful analysis. After applying this filter, we selected 155 projects for our study. To observe the prevalence of community smells based on the community size, we categorized the projects into small, medium, and large projects following Tamburri et al. \cite{tamburri2019exploring}. Table~\ref{tab:project_breakdown} shows the breakdown of the project types.

\begin{figure*}[htb]
    \centering
    \captionsetup[subfloat]{labelfont=large,textfont=Large}
    \resizebox{\textwidth}{!}{
            \subfloat[Small projects]{
            \begin{tikzpicture}
                 \tikzstyle{every node}=[font=\large]
                    \begin{axis}
                        [
                            boxplot/draw direction=y,
                            xtick= {1,2,3,4,5,6,7,8,9,10},
                            xticklabels={OSE,BCE,PDE,SV,OS,SD,RS,TF,UI,TC},
                            x tick label style={align=center,  font=\large},
                            y tick label style={font=\large},
                            x=1cm
                        ]
                        \addplot+   [ boxplot prepared={lower whisker=0,lower quartile=0,median=1,upper quartile=3,upper whisker=7.5}] coordinates {};
                        \addplot+   [only marks,mark=*,mark options={scale=0.9}] coordinates {(1, 2.59)};   
                        \addplot+   [ boxplot prepared={lower whisker=1,lower quartile=8.25,median=15.5,upper quartile=28.75,upper whisker=59.50}] coordinates {};
                        \addplot+   [only marks,mark=*,mark options={scale=0.9}] coordinates {(2, 34.11)}; 
                        \addplot+   [boxplot prepared={lower whisker=1.0, lower quartile=9.00, median=16, upper quartile=29.50, upper whisker=60.25}] coordinates {};
                        \addplot+   [only marks,mark=*,mark options={scale=0.9}] coordinates {(3, 35.66)};
                        \addplot+   [boxplot prepared={lower whisker=0.0, lower quartile=1.00, median=3, upper quartile=5, upper whisker=11}] coordinates {};
                        \addplot+   [only marks,mark=*,mark options={scale=0.9}] coordinates {(4, 5.9)};
                        \addplot+   [boxplot prepared={lower whisker=0.0, lower quartile=0.00, median=0, upper quartile=0, upper whisker=0}] coordinates {};
                        \addplot+   [only marks,mark=*,mark options={scale=0.9}] coordinates {(5, 0.3)};
                        \addplot+   [boxplot prepared={lower whisker=0.0, lower quartile=0.00, median=1.0, upper quartile=2.0, upper whisker=5.0}] coordinates {};
                        \addplot+   [only marks,mark=*,mark options={scale=0.9}] coordinates {(6, 1.72)};
                        \addplot+   [boxplot prepared={lower whisker=0.0, lower quartile=5.00, median=21.0, upper quartile=20.0, upper whisker=42.5}] coordinates {};
                        \addplot+   [only marks,mark=*,mark options={scale=0.9}] coordinates {(7,35.51)};
                        \addplot+   [boxplot prepared={lower whisker=1.0, lower quartile=9.00, median=16.0, upper quartile=29.5, upper whisker=60.25}] coordinates {};
                        \addplot+   [only marks,mark=*,mark options={scale=0.9}] coordinates {(8,35.51)};
                        \addplot+   [boxplot prepared={lower whisker=0.0, lower quartile=0.00, median=1.0, upper quartile=3.0, upper whisker=7.5}] coordinates {};
                        \addplot+   [only marks,mark=*,mark options={scale=0.9}] coordinates {(9,5.07)};
                        \addplot+   [boxplot prepared={lower whisker=1.0, lower quartile=9.00, median=16.0, upper quartile=29.5, upper whisker=60.25}] coordinates {};
                        \addplot+   [only marks,mark=*,mark options={scale=0.9}] coordinates {(10,35.7)};

                    \end{axis}
                \end{tikzpicture}
            }
            \subfloat[Medium projects]{
            \begin{tikzpicture}
                 \tikzstyle{every node}=[font=\small]
                    \begin{axis}
                        [
                            boxplot/draw direction=y,
                            xtick= {1,2,3,4,5,6,7,8,9,10},
                            xticklabels={OSE,BCE,PDE,SV,OS,SD,RS,TF,UI,TC},
                            x tick label style={align=center,  font=\large},
                            y tick label style={font=\large},
                            x=1cm
                        ]
                        \addplot+ [boxplot prepared={lower whisker=0.0, lower quartile=2.0, median=5.0, upper quartile=7.0, upper whisker=14.5}] coordinates {};
                        \addplot+ [only marks, mark=*, mark options={scale=0.9}] coordinates {(1, 5.87)};               
                        \addplot+ [boxplot prepared={lower whisker=1.0, lower quartile=13.0, median=32.0, upper quartile=55.0, upper whisker=118.0}] coordinates {};
                        \addplot+ [only marks, mark=*, mark options={scale=0.9}] coordinates {(2, 38.07)};                
                        \addplot+ [boxplot prepared={lower whisker=1.0, lower quartile=19.0, median=34.0, upper quartile=54.0, upper whisker=106.5}] coordinates {};
                        \addplot+ [only marks, mark=*, mark options={scale=0.9}] coordinates {(3, 40.40)};                
                        \addplot+ [boxplot prepared={lower whisker=0.0, lower quartile=4.0, median=6.0, upper quartile=10.0, upper whisker=19.0}] coordinates {};
                        \addplot+ [only marks, mark=*, mark options={scale=0.9}] coordinates {(4, 7.56)};                
                        \addplot+ [boxplot prepared={lower whisker=0.0, lower quartile=0.0, median=1.0, upper quartile=2.0, upper whisker=5.0}] coordinates {};
                        \addplot+ [only marks, mark=*, mark options={scale=0.9}] coordinates {(5, 1.82)};               
                        \addplot+ [boxplot prepared={lower whisker=0.0, lower quartile=0.0, median=1.0, upper quartile=3.0, upper whisker=7.5}] coordinates {};
                        \addplot+ [only marks, mark=*, mark options={scale=0.9}] coordinates {(6, 2.42)};               
                        \addplot+ [boxplot prepared={lower whisker=1.0, lower quartile=12.0, median=25.0, upper quartile=47.0, upper whisker=99.5}] coordinates {};
                        \addplot+ [only marks, mark=*, mark options={scale=0.9}] coordinates {(7, 32.51)};               
                        \addplot+ [boxplot prepared={lower whisker=2.0, lower quartile=16.0, median=30.0, upper quartile=55.0, upper whisker=113.5}] coordinates {};
                        \addplot+ [only marks, mark=*, mark options={scale=0.9}] coordinates {(8, 38.82)};               
                        \addplot+ [boxplot prepared={lower whisker=0.0, lower quartile=0.0, median=1.0, upper quartile=3.0, upper whisker=7.5}] coordinates {};
                        \addplot+ [only marks, mark=*, mark options={scale=0.9}] coordinates {(9, 3.67)};                
                        \addplot+ [boxplot prepared={lower whisker=0.0, lower quartile=19.0, median=34.0, upper quartile=55.0, upper whisker=109.0}] coordinates {};
                        \addplot+ [only marks, mark=*, mark options={scale=0.9}] coordinates {(10, 40.47)};

                    \end{axis}
                \end{tikzpicture}
            }
            \subfloat[Large projects]{
            \begin{tikzpicture}
                 \tikzstyle{every node}=[font=\small]
                    \begin{axis}
                        [
                            boxplot/draw direction=y,
                            xtick= {1,2,3,4,5,6,7,8,9,10},
                            xticklabels={{OSE},{BCE},{PDE},{SV},{OS},{SD},{RS},{TF},{UI},{TC}},
                            x tick label style={align=center,  font=\large},
                            y tick label style={font=\large},
                            x=1cm
                        ]
                        \addplot+ [boxplot prepared={lower whisker=0.0, lower quartile=2.0, median=5.0, upper quartile=13.0, upper whisker=25.0}] coordinates {};
                        \addplot+ [only marks, mark=*, mark options={scale=0.9}] coordinates {(1, 7.32)};                
                        \addplot+ [boxplot prepared={lower whisker=5.0, lower quartile=18.5, median=32.0, upper quartile=45.0, upper whisker=84.75}] coordinates {};
                        \addplot+ [only marks, mark=*, mark options={scale=0.9}] coordinates {(2, 37.48)};                
                        \addplot+ [boxplot prepared={lower whisker=10.0, lower quartile=29.5, median=43.0, upper quartile=55.0, upper whisker=93.25}] coordinates {};
                        \addplot+ [only marks, mark=*, mark options={scale=0.9}] coordinates {(3, 47.94)};                
                        \addplot+ [boxplot prepared={lower whisker=1.0, lower quartile=7.0, median=12.0, upper quartile=16.0, upper whisker=29.5}] coordinates {};
                        \addplot+ [only marks, mark=*, mark options={scale=0.9}] coordinates {(4, 12.65)};                
                        \addplot+ [boxplot prepared={lower whisker=0.0, lower quartile=1.0, median=4.0, upper quartile=7.5, upper whisker=17.25}] coordinates {};
                        \addplot+ [only marks, mark=*, mark options={scale=0.9}] coordinates {(5, 5.61)};                
                        \addplot+ [boxplot prepared={lower whisker=0.0, lower quartile=0.0, median=0.0, upper quartile=2.0, upper whisker=5.0}] coordinates {};
                        \addplot+ [only marks, mark=*, mark options={scale=0.9}] coordinates {(6, 1.42)};                
                        \addplot+ [boxplot prepared={lower whisker=5.0, lower quartile=19.0, median=33.0, upper quartile=45.0, upper whisker=84.0}] coordinates {};
                        \addplot+ [only marks, mark=*, mark options={scale=0.9}] coordinates {(7, 36.61)};                
                        \addplot+ [boxplot prepared={lower whisker=10.0, lower quartile=24.0, median=35.0, upper quartile=46.0, upper whisker=79.0}] coordinates {};
                        \addplot+ [only marks, mark=*, mark options={scale=0.9}] coordinates {(8, 40.03)};                
                        \addplot+ [boxplot prepared={lower whisker=0.0, lower quartile=0.0, median=4.0, upper quartile=6.0, upper whisker=15.0}] coordinates {};
                        \addplot+ [only marks, mark=*, mark options={scale=0.9}] coordinates {(9, 5.45)};                
                        \addplot+ [boxplot prepared={lower whisker=10.0, lower quartile=29.0, median=43.0, upper quartile=55.0, upper whisker=94.0}] coordinates {};
                        \addplot+ [only marks, mark=*, mark options={scale=0.9}] coordinates {(10, 47.35)};

                    \end{axis}
                \end{tikzpicture}
            }
        }
    
    \caption{Quartile analysis of community smells by project size  ($\bullet$ represents the mean value).}
    \label{fig:community-smells}
\end{figure*}
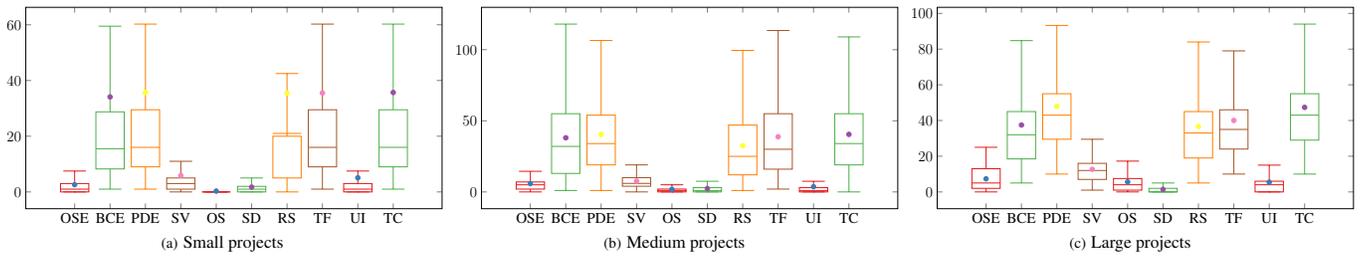

\begin{table*}[ht]
\centering
\caption{Quartile statistics of SATD-related community smell occurrences by project size}
\label{tab:quartile-satd-correlation}
\resizebox{0.6\textwidth}{!}{
\begin{tabular}{llcccccccccc}
\hline
\textbf{Size} & \textbf{Metric} & \textbf{OSE} & \textbf{BCE} & \textbf{PDE} & \textbf{SV} & \textbf{OS} & \textbf{SD} & \textbf{RS} & \textbf{TF} & \textbf{UI} & \textbf{TC} \\
\hline
\multirow{5}{*}{Small}
& Median       & 1.00  & 15.50 & 16.00 & 3.00  & 0.00  & 1.00  & 11.00 & 16.00 & 1.00  & 16.00 \\
& Average      & 2.59  & 34.11 & 35.66 & 5.90  & 0.30  & 1.72  & 26.98 & 35.51 & 5.07  & 35.70 \\
& Min          & 0.00  & 1.00  & 1.00  & 0.00  & 0.00  & 0.00  & 0.00  & 1.00  & 0.00  & 1.00  \\
& Max          & 7.50  & 59.50 & 60.25 & 11.00 & 0.00  & 5.00  & 42.50 & 60.25 & 7.50  & 60.25 \\
& Std. Dev     & 6.05  & 104.18&110.54 &15.64  & 1.29  & 5.13  &102.79 &110.56 &24.75  &110.53 \\
\hline
\multirow{5}{*}{Medium}
& Median       & 5.00  & 32.00 & 35.50 & 6.00  & 1.00  & 1.00  & 25.50 & 30.50 & 1.00  & 35.50 \\
& Average      & 5.85  & 38.06 & 40.40 & 7.54  & 1.81  & 2.41  & 32.51 & 38.82 & 3.66  & 40.47 \\
& Min          & 0.00  & 1.00  & 1.00  & 0.00  & 0.00  & 0.00  & 1.00  & 2.00  & 0.00  & 0.00  \\
& Max          &14.50  &116.63 &105.88 &18.38  & 4.59  & 7.50  &96.50  &112.25 & 7.80  &108.38 \\
& Std. Dev     & 6.50  & 28.76 & 28.09 & 6.62  & 3.98  & 4.56  &26.15  &28.59  & 6.58  & 28.11 \\
\hline
\multirow{5}{*}{Large}
& Median       & 5.50  & 31.50 & 42.50 & 12.00 & 4.00  & 0.50  & 32.00 & 34.00 & 4.00  & 42.50 \\
& Average      & 7.43  & 34.63 & 45.33 & 12.90 & 5.77  & 1.47  & 33.83 & 37.27 & 5.63  & 44.73 \\
& Min          & 0.00  & 5.00  & 10.00 & 1.00  & 0.00  & 0.00  & 5.00  &10.00  & 0.00  &10.00  \\
& Max          &25.00  &80.25  &93.63  &29.13  &17.88  & 5.00  &75.25  &75.25  &14.63  &94.00  \\
& Std. Dev     & 6.45  & 23.16 & 25.05 & 8.28  & 6.37  & 2.70  &22.89  &23.90  & 7.93  &25.73  \\
\hline
\end{tabular}
}
\end{table*}

\begin{table}[htb]
    \centering
    \caption{\# of ML projects affected by community smells}
    \large
    \resizebox{0.8\linewidth}{!}{
    \begin{tabular}{p{1.75cm}cccccccccc}
    \hline
    Project type &   OSE & BCE & PDE & SV  &	OS &	SD &	RS &	TF &	UI &	TC\\
    \hline
        Small & 53 & 	81 & 	81 & 	69 & 	11 & 	46 & 	80 & 	81 & 	44 & 	81 \\
        Medium & 39 & 	44 & 	44 & 	42 & 	23 & 	29 & 	44 & 	44 & 	26 & 	43\\
        Large & 25 & 	30	 & 30 & 	30 & 	25 & 	14 & 	30 & 	30 & 	21 & 	30\\
        Overall &  117 & 	155 & 	155 & 	141 & 	59 & 	89 & 	154	 & 155 & 	91 & 	154 \\
    \hline
    \end{tabular}
    }
    \label{tab:community-dimension}
     \vspace{-0.5em}
\end{table}
\begin{table}[h]
    \centering
    \Large
    \caption{Definitions and examples of six SATD types \cite{obrien202223}}
    \resizebox{0.9\linewidth}{!}{
            \begin{tabular}{p{3cm}|p{9cm}|p{6cm}}
            \hline
            \begin{tabular}{@{}c@{}}SATD\\Types\end{tabular}& Descriptions  &\begin{tabular}{@{}c@{}}Example\\Comment\end{tabular} \\ \hline
            Requirement & Requirement debt can either be functional (incomplete features) or non-functional (falls performance, memory, or security standards). &  TODO: handle channel modalities later \\\hline
            Code    &  Poor coding practices reduce code legibility, making it hard to understand and maintain. & TODO: This next code is dense and confusing. Clean up at some point. \\\hline
            Monitoring and Testing (M\&T)    &   Issues in implementing testing or monitoring components. &  XXX: should we rather test if instance of estimator? \\\hline
            Defect    &   Detected system defects that need resolution    &  TODO this will fail if a parameter cant handle size=(N;) \\\hline
            Design    &   Violations of good design practices that hinder adaptability to changing business needs &  TODO maybe improve this so it doesn't use a global. \\\hline
            Documentation    &   Insufficient documentation within the software system.  & TODO update doc above.  \\
            \hline
        \end{tabular}
    }
    \label{tab:SATD_types}
    \vspace{-1em}
\end{table}
\subsection{Analyzing the Prevalence of Community Smells}
\label{methodology-csDetector}
In order to detect community smells for every release of each project, we utilized the csDetector tool \cite{almarimi2021csdetector}. This sophisticated tool utilizes an ML-based detection method that provides insights from numerous instances of suboptimal community development practices, facilitating automated identification of similar issues within the community. The csDetector tool is well-known in the software engineering community \cite{almarimi2023improving,neugebauer2024characteristic,caballero2023community,linaaker2022characterize,jabrayilzade2022bus}. There are some other tools mentioned in the existing studies \cite{tamburri2019discovering,voria2022community,della2024novel,8651329}. However, in our study, we use csDetector for several key reasons: (1) the tool offers a comprehensive framework for detecting community smells by integrating diverse data sources, including social network graph analysis (covering commits, pull requests, and issue report discussions), sentiment analysis, and truck factor metrics, (2) csDetector exhibits high performance, achieving an average F1 score of 84\% across different types of community smells, based on an analysis of 143 open-source projects and (3) csDetector can detect ten types of community smells and it detects the smells based on 52 socio-technical metrics whereas the other state-of-the-art tools can detect four types of community smells \cite{8651329} or identify development communities based on six factors \cite{tamburri2019discovering}.  

Next, we introduced two parameters in the csDetector tool, \textit{start\_date} and \textit{end\_date}, to enable release-based analysis of the repositories. For each project, the tool iterates through the repository from the date of one release to the next, detecting community smells within that specific time frame. Since csDetector relies on ML-based detection methods, the results may vary slightly across runs due to inherent non-determinism in the model \cite{faria2017non}. To ensure the reliability of the outputs, we conducted a robustness check: for a randomly selected sample of 10 projects, we executed the tool five times per release and manually examined the detected smells. The first two authors of this study independently reviewed the outputs. To improve transparency and address concerns about the depth of our manual analysis, we developed a rubric based on prior literature to guide our classification and ensure structured, repeatable assessment.
In the event of a conflict regarding the presence of a community smell, where there is disagreement about its existence, the authors seek to resolve it through open discussion. This approach aims to reach a consensus on the key symptoms of the potential smell. Our validation process revealed that in one out of the ten projects, the detection of the Toxic Communication (TC) smell did not reach full agreement for some releases. Consequently, we excluded that smell from the release's detected smells and re-evaluated the detection of this specific smell across all other projects. Ultimately, the authors achieved full agreement with the tool's results, ensuring the reliability and accuracy of the community smell detections in our study.
Next, following the methodology of Tamburri et al.\cite{8651329}, we conducted a quartile analysis to examine the distribution and prevalence of community smells across ML-based projects.

\subsection{Investigating the Correlations between Community Smells and SATD}
We investigated the correlation between community smells and SATD occurrences in ML projects using Spearman’s rank correlation \cite{zar2005spearman}, as the data was not normally distributed~\cite{prion2014making}. Spearman’s correlation measures the strength and direction of monotonic relationships between variables. To detect a moderate effect size ($\rho$ = 0.3) with $\alpha$ = 0.05 and power = 0.8, a minimum of 82 observations is typically required~\cite{bujang2016sample}. We selected a stratified sample of 36 projects (19 small, 10 medium, 7 large), preserving the original size distribution of the dataset (52.6\% small, 28.2\% medium, 19.2\% large)~\cite{pontillo2024machine}. These projects collectively contain over 800 releases, offering more than enough data points to meet the power requirements and support reliable correlation analysis.
After identifying the sampled projects, we proceeded to extract SATD-related comments to analyze their relationship with community smells. We collected source code comments from each project's release history by cloning their GitHub repositories and checking out each tagged release. We targeted source files with common extensions (.py, .java, .js, .cpp, .c), using predefined patterns to extract single-line comments. These comments were aggregated per release and saved in structured JSON files.

To identify SATD instances within the extracted comments, we employed the MT-BERT-SATD model proposed by Gu et al.\cite{gu2024self}, which is based on multi-task learning using BERT. The model achieved an average F1-score of 0.712 (ranging from 0.625 to 0.859), outperforming previous approaches by 4.6\% to 30.1\%. The model has demonstrated effective generalization across programming languages and software artifacts. We applied this model to the collected comment sets and stored the predictions for each project and release. For every release, we recorded both the number of detected SATD instances and the number of community smells. The complete results are available in the online appendix \cite{replication_package}. Finally, we applied Spearman’s rank correlation independently for each project to examine whether a statistically significant correlation exists between these two factors.

\subsection{Exploring the association between SATD types and community smells}
To further examine the relationship between community smells and SATD, we analyzed which types of SATD are more closely associated with specific community smells. Therefore, we classified the detected SATD comments using a Large Language Model (LLM). Sheikhaei et al.~\cite{sheikhaei2024empirical} evaluated the effectiveness of LLMs in both identifying and classifying SATD comments, and found that an ensemble approach based on the Flan-T5-XL \cite{flan-t5} model outperformed other models in terms of classification accuracy. Following their methodology, we used the same model and settings to categorize SATD instances in our dataset.
In terms of SATD taxonomy, prior work by O'Brien et al.\cite{obrien202223}, introduced \textit{six} general categories of SATD, i.e., Requirement, Code, M\&T, Defect, Design, and Documentation, as well as 23 domain-specific subcategories for ML-related projects. However, since Sheikhaei et al.'s \cite{sheikhaei2024empirical} work focused on general SATD classification and we aimed to explore the broader associations across project types, we adopted only the six general SATD categories in our study. Table~\ref{tab:SATD_types} summarizes the description of six types of SATD. The classified SATD types were stored alongside the community smell detection results for each project release. Full results are available in the online appendix \cite{replication_package}.

Next, to analyze the association between specific SATD types and community smells, we applied the Point-Biserial Correlation test~\cite{kornbrot2014point,tate1954correlation}. This statistical method is appropriate when one variable is binary and the other is continuous. In our case, the presence or absence of each of the ten community smells types serves as the binary variable, while the number of instances for each of the six SATD types serves as the continuous variable. This allowed us to assess the strength of association between individual SATD types and specific community smells across project releases.

\subsection{Analyzing trends of community smells and SATD over releases} We employed the Mann-Kendall test \cite{hamed1998modified} to assess whether community smells and SATD instances exhibit upward, downward, or no trends over project releases. The null hypothesis assumes no trend, while the alternative suggests a monotonic trend. The test’s coefficient indicates the direction and strength of the trend. By applying this test to each project, we examined whether community smells and SATD evolve in parallel over time. A consistent trend in both metrics may suggest a potential relationship between them.

\section{Findings}
\subsection{Answering RQ1: Prevalence of community smells in ML projects}

Fig.~\ref{fig:community-smells} depicts the box plots reporting the distribution of the community smells analyzed over the 155 ML projects in our dataset. These plots reflect the overall distributions by aggregating community smell occurrences over all releases in the three categories. Next, Table~\ref{tab:quartile-satd-correlation} reports the median, average, minimum, maximum, and standard deviation observed in each of the releases for all the projects. 
As shown in Fig.~\ref{fig:community-smells}, five community smells, Black Cloud Effect (BCE), Prima-donna Effect (PDE), Radio Silence (RS), Truck Factor Smell (TFS), and Toxic Communication (TC), are the most prevalent across all project sizes, with consistently high median and mean values. In small projects, the average occurrences of BCE, PDE, TFS, and TC range from 34-36, though high standard deviations (above 100) suggest that some teams face severe socio-technical issues despite lower overall averages.  
However, these smells intensified in medium projects, with PDE and TC reaching median values of 35.5, BCE at 32.0, and TFS at 30.5. RS also showed increased presence, with a median of 25.5. Interestingly, large projects followed a similar trend, exhibiting medians of 42.5 for both PDE and TC, 31.5 for BCE, 34.0 for TFS, and 32.0 for RS. The higher differences of PDE and TC in large projects compared to small projects suggest that smells related to knowledge centralization and hostile communication become significantly more prominent as projects scale.
Furthermore, Sharing Villainy (SV) and Organizational Silo Effect (OSE) show a steady increase in median values from small to large projects, with SV rising from 3.0 to 12.0 and OSE from 1.0 to 5.5. This reflects a trend where growing team size and complexity lead to less efficient knowledge sharing and more isolated collaboration. In contrast, Organizational Silo Effect (OSE), Solution Deficiency (SD), and Unhealthy Interaction (UI) were less prevalent overall but displayed unique distribution patterns worth further investigation. 

Next, from Table~\ref{tab:community-dimension}, we observed that the five most frequent community smells are present in almost all of the projects analyzed. Each of these smells was detected in at least 154 out of 155 projects. This widespread presence indicates that the occurrence of these smells is not solely dependent on community size. However, the pattern is notably different for the Organizational Skirmish (OS) smell. Only 11 projects with fewer than 50 contributors exhibited OS, and the number of affected projects gradually increases with larger community sizes. This trend suggests that the presence of OS may be influenced by the size of the community, indicating a potential correlation between contributor count and OS occurrences.

\tcbset{colback=blue!7!white,colframe=gray!50!black} 
\begin{tcolorbox}[leftrule=3mm, boxrule=1pt] 
\textbf{Summary for RQ\textsubscript{1}:} Community smells are highly prevalent in ML projects but not uniformly distributed. BCE, PDE, RS, TFS and TC are found regularly, while the presence of OS instances may depend on the number of contributors within a community. 
\end{tcolorbox}

\subsection{Answering RQ2: Correlation Between Community Smells and SATD Types }

The previous section showed the prevalence of community smells in ML projects. This section analyzes whether the community smells help in the emergence of SATD instances, i.e., whether community smells have any correlation with SATD instances.
\begin{table}[htb]
    \centering
    \caption{Spearman Rank Correlation Between Community Smells and SATD Per Project}
    \normalsize
    \resizebox{0.9\linewidth}{!}{
    \begin{tabular}{p{1.5cm}|c|l|c|c|c}
        \hline
Community dimension & ID   & Project          &   $\rho$  &   $p-value$   &   Significance    \\
        \hline
        small   &P1        & bigartm                   &   0.73    &   0.0002      &   $***$    \\
                &P2        & coco-annotator                   &   0.53    &   0.005       &   $**$     \\
                &P3        & PyKEEN                   &   0.57    &   0.009       &   $**$     \\   
                &P4        & POT                   &   0.68    &   0.0000      &   $***$    \\
                &P5         & OpenVINO-model-server                  &   0.44    &   0.03        &  $*$       \\  
                &P6         & dm-control                   &  0.47     &   0.01        &   $*$      \\
                &P7         & EconML                  &   0.41    &   0.02        &   $*$      \\
                &P8         & interpret                 &   0.33    &   0.02        &   $*$      \\
        \hline
        medium  &P9          & pyro                 &   0.71    &   0.0000      &   $***$    \\
                &P10         & ludwig                 &   0.55    &   0.0000      &   $***$    \\
                &P11         & hdbscan                 &   0.44    &   0.0008      &   $***$    \\
                &P12          & compromise               &   0.44    &   0.0000      &   $***$    \\
                &P13          & nlp.js               &   0.34    &   0.02        &   $*$      \\
                &P14         & scanpy                &   0.27    &   0.01        &   $*$      \\
                &P15          & MMdnn                &   0.70    &   0.02        &   $*$      \\
                &P16          & deepchem                &   0.58    &   0.009       &   $**$     \\
        \hline
        large   &P17           & tesseract                &   0.48    &   0.006       &   $**$     \\
                &P18           & incubator-mxnet               &   0.55    &   0.0015      &   $**$     \\    
                &P19           & shap               &   0.28    &   0.04        &   $*$      \\
                &P20           & mne-python              &   0.28    &   0.04        &   $*$      \\
                &P21           & OpenNMT-py              &   0.32    &   0.03        &   $*$      \\
                &P22           & kornia              &   0.33    &   0.04        &   $*$      \\
        \hline
        \multicolumn{5}{l}{\textit{Note:} $^*p<0.05$, $^{**}p<0.01$, $^{***}p<0.001$}
    \end{tabular}
    }
    \label{tab:all-project-smell}
    \vspace{-2em}
\end{table}
\begin{table*}[htb]
    \centering
    \caption{Spearman Rank Correlation Between Individual Community Smells and SATD Per Project}
    \resizebox{0.7\textwidth}{!}{
        \begin{tabular}{l|l|c|c|c|c|c|c|c|c|c|c}
        \hline
            & Project                  & OSE      & BCE      & PDE      & SV       & OS       & SD       & RS       & TF       & UI       & TC       \\
        \hline
        \multirow{8}{*}{Small} 
            & P1         & 0.613$^{***}$ & 0.368    & --       & 0.346    & --       & 0.354 & 0.368    & 0.368    & 0.198    & --       \\
            & P2  & 0.448$^*$ & --       & --       & 0.358    & --       & 0.358    &  0.456$^*$ & -0.248   & --       & --       \\
            & P3          & 0.352    & 0.02     & --       & 0.566$^{**}$ & --       & --       &  --       & --       & --       & --       \\
            & P4             & 0.615$^{***}$ & -0.129   & --       & 0.501$^{**}$ & --       & 0.341    & 0.516$^{**}$ & --       & -0.033   & --       \\
            & P5    & --       & 0.122    & --       & 0.137    & --       & --       & 0.286    & 0.274    & 0.187    & --       \\
            & P6              & -0.085   & --       & 0.451$^*$ & 0.17     & --       & --       & --       & --       & --       & --       \\
            & P7            & 0.205    & --       & --       & 0.481$^{**}$ & --       & --    &  -0.092   & --       & --       & --       \\
            & P8                & 0.025    & -0.05    & --       & 0.216    & --       & --       & 0.174    & --       & 0.166    & --       \\        
             
        \hline
        \multirow{7}{*}{Medium} 
            & P9                     & 0.481$^{**}$ & 0.287    & --       & 0.446$^{**}$ & --       & 0.151   & 0.275    & 0.356$^*$ & --       & --       \\
            & P10                   & 0.011    & 0.401$^{**}$ & --       & -0.315$^*$ & -0.38$^{**}$ & --  & 0.498$^{***}$ & 0.353$^*$ & -0.194   & --       \\
            & P11                  & 0.297$^*$ & --       & --       & 0.446$^{***}$ & --       &  0.322$^*$ & 0.316$^*$ & -0.148   & --       & --       \\
            & P12               & 0.123    & -0.192   & --       & 0.262$^*$ & 0.108    & 0.137    &  0.383$^{***}$ & -0.108   & 0.175    & --       \\
            & P13                  & 0.116    & -0.116   & --       & 0.167    & --       & 0.116    &  0.296    & --       & 0.116    & --       \\
            & P14                   & 0.201    & --       & --       & 0.35$^{***}$ & --       & --      & -0.118   & --       & --       & --       \\
            & P15                    & 0.574    & 0.43     & --       & 0.574    & 0.0      & 0.382    & 0.344    & 0.462    & -0.205   & --       \\
            & P16                 & --       & -0.094   & --       & 0.02     & 0.25     & 0.391    & 0.166    & 0.311    & 0.011    & --       \\
                
        \hline
        \multirow{7}{*}{Large} 
            & P17                 & 0.402$^*$ & 0.565$^{***}$ & 0.604$^{***}$ & -0.224   & --  & -0.455$^{**}$ & 0.689$^{***}$ & 0.131    & -0.578$^{***}$ & 0.616$^{***}$ \\
            & P18          & 0.148    & 0.417$^*$ & --       & 0.008    & 0.387$^*$ & --  & 0.257    & 0.322    & 0.017    & 0.291    \\
            & P19                      & 0.107    & -0.05    & --       & 0.282$^*$ & -0.053   & 0.166    & 0.177    & -0.307$^*$ & 0.109    & --       \\
            & P20               & 0.387$^{**}$ & 0.219    & --       & -0.278$^*$ & 0.019    & --   & 0.426$^{***}$ & 0.16     & -0.538$^{***}$ & --       \\
            & P21               & 0.126    & 0.25     & --       & 0.161    & -0.25    & -0.036   &  0.071    & 0.283$^*$ & --       & --       \\
            & P22                    & 0.316$^*$ & -0.08    & --       & 0.168    & 0.412$^{**}$ & --  & 0.307    & -0.14    & -0.236   & --       \\
            
        \hline
        \multicolumn{12}{l}{\textit{Note:} $^*p<0.05$, $^{**}p<0.01$, $^{***}p<0.001$; -- indicates insufficient data for correlation.}
        \end{tabular}
    }
    \label{tab:spearman_correlations}
\end{table*}
\begin{figure*}[htb]
    \centering
    \captionsetup[subfloat]{labelfont=large,textfont=Large}
    \caption{Correlation between individual community smells and SATD types across all projects}
    \resizebox{\textwidth}{!}{
        \subfloat[Small projects]{
            \includegraphics[width=0.7\linewidth]{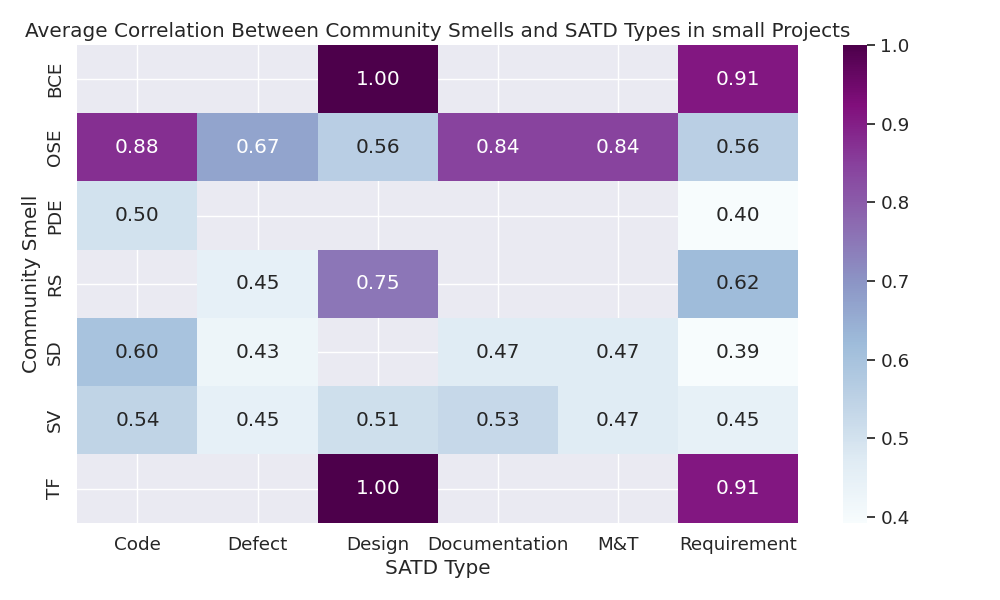}
            \label{small-project}
        } 
        \subfloat[Medium projects]{
            \includegraphics[width=0.7\linewidth]{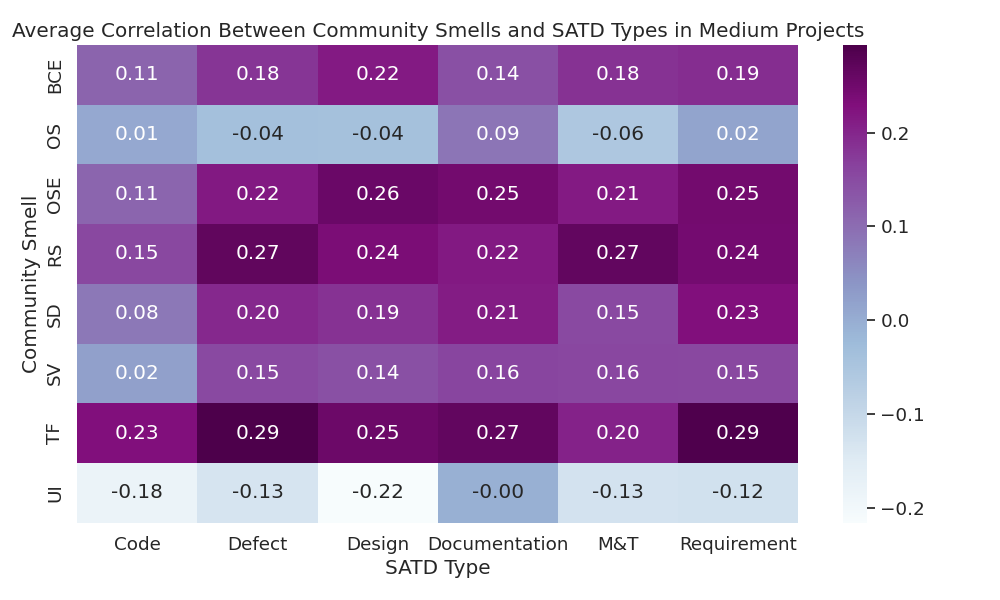}
            \label{medium-project}
        }
        \subfloat[Large projects]{
            \includegraphics[width=0.7\linewidth]{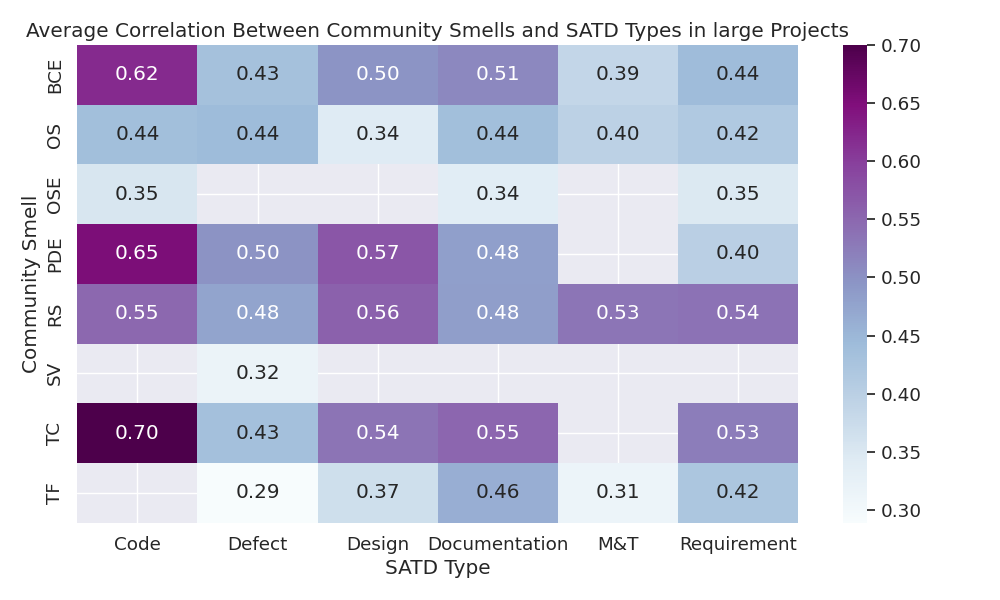}
            \label{large-project}
        }
    }
    \label{fig:satd-type}
    \vspace{-1em}
\end{figure*}
As shown in Table~\ref{tab:all-project-smell}, we found significant correlations with SATD in 8 out of 19 cases among the small projects; for medium projects, 8 out of 10 exhibited significant relationships; and in large projects, 6 out of 7 demonstrated significant associations. Overall, our Spearman rank correlation analysis indicates that releases with a higher number of community smells tend to exhibit an increased number of SATD instances. 
Based on this finding, we further analyzed the relationship between individual community smells and the number of SATD instances per release across 23 ML projects. Table~\ref{tab:spearman_correlations} summarizes the overall result. 
As shown in the Table, Radio Silence (RS) emerges as a consistently significant factor. Small projects such as \textit{P2} and \textit{P4}, medium projects such as \textit{P10} and \textit{P12}, and large projects such as \textit{P17} and \textit{P20} all exhibit strong, positive correlations between RS and SATD. These findings suggest that procedural bottlenecks and delayed communication may contribute to the accumulation of SATD as necessary changes are de-prioritized in overly formal development environments. 
Sharing Villainy (SV) consistently correlates with SATD across small (e.g., \textit{P3}, \textit{P7}), medium (\textit{P9}, \textit{P14}), and large (\textit{P19}) projects, suggesting that outdated or overly specialized solutions contribute to technical debt. In fast-evolving ML systems, such knowledge gaps delay debt resolution for the entire team. ML systems evolve rapidly, so gaps or outdated knowledge slow down technical debt fixes for everyone.
Organizational Silo Effect (OSE) consistently correlates with SATD across project sizes, small (\textit{P1}, \textit{P4}), medium (\textit{P9}, \textit{P11}), and large (\textit{P17}, \textit{P20}). Silos fragment collaboration  hindering coordinated debt management. As teams grow, OSE’s prevalence rises, further reinforcing the idea that fractured group structures become more detrimental with increasing project size. In ML projects such silos may prevent timely resolution of design or implementation trade-offs, thereby contributing to sustained or growing SATD levels. Together, RS, SV, and OSE emerge as key socio-technical challenges tied to SATD in ML projects.

However, despite overall similarities, each project size shows unique patterns between communit smells and SATD. Small projects often have few or no significant correlations, but when present, the effects can be strong. For example, P4 shows strong associations with OSE, SV and RS, whereas \textit{P5} and \textit{P6} lack enough data for meaningful analysis.
Interestingly, medium projects reveal more complex and sometimes conflicting relationships. Projects such as \textit{P10} demonstrate both positive (OSE, BCE, RS, TF) and negative (SV, OS, UI) correlations within the same community, underscoring that a moderate increase in contributors can yield a mixture of organizational and communication patterns. 
%
In contrast, large projects show the highest frequency and diversity of significant results. Projects such as \textit{P17} and \textit{P20} display multiple strong correlations, both positive and negative. Notably, Unhealthy Interaction (UI) is more frequently associated with a negative correlation with SATD in large projects. Conversely, strong positive correlations with the Prima-donna Effect (PDE) and Toxic Communication (TC) are more prominent in large-scale settings, suggesting that issues such as power concentration and hostility become increasingly impactful as contributor communities grow.

\tcbset{colback=blue!7!white,colframe=gray!50!black} 
\begin{tcolorbox}[leftrule=3mm, boxrule=1pt] 
\textbf{Summary for RQ\textsubscript{2}:} Community smells consistently relate to SATD across ML projects, with RS, SV, and OSE being particularly prevalent. However, the influence of UI, PDE, and TC varies by project size and complexity.
\end{tcolorbox}

\subsection{Answering RQ3: Correlation Between Community Smells and SATD Across ML Projects}
We found that there are significant correlations between community smells and SATD instances across all the project sizes. In this section, we wanted to analyze how these community smells correlate with specific types of SATD.
As shown in Fig.~\ref{small-project}, we observed that in small projects, there exists strong and statistically significant positive correlations between specific community smells and particular types of SATD. Notably, Organization Silo Effect (OSE) shows strong correlations with multiple SATD categories, e.g., Code, Documentation, Testing, and Defect. This result implies that isolated communication structures are linked with both implementation and process-level debt. Similarly, Black Cloud Effect (BCE) and Truck Factor (TF) exhibit perfect correlations with Design debt and high correlations with Requirement debt, indicating that over-reliance on specific individuals contributes to architectural and design-related technical challenges. Conversely, smells like Sharing Villainy (SV) and Solution Defiance (SD) show moderate or weak associations across SATD types, while Radio Silence (RS) demonstrates a moderate link to Design and Requirement debt.  

Compared to small projects, the associations are generally weaker but reveal meaningful patterns in Fig.~\ref{medium-project}. Notably, Truck Factor (TF) exhibits the strongest and most consistent correlations across all SATD types, particularly with Defect, Documentation, and Requirement debt. Similarly, Organizational Silo Effect (OSE) and Radio Silence (RS) show moderate correlations with several SATD categories especially in Desgin, Defect and Requirement debt. In contrast, smells such as Unhealthy Interactions (UI) and Organizational Skirmish (OS) exhibit weak or even negative correlations across most SATD types. 
Notably, for large repositories, the results in Fig.~\ref{large-project} demonstrate strong and more widespread associations compared to small and medium-sized projects, which highlights how socio-technical challenges scale with team size. Toxic Communications (TC) and Prima-donna Effect (PDE) exhibit particularly strong correlations across multiple SATD categories. TC shows the highest correlations with Code debt and notable associations with Documentation, Design, and Requirement debts. Similarly, PDE is strongly related to Code, Design, and Defect debt. Radio Silence (RS) also emerges as a prominent factor, moderately correlated with five SATD types, most notably Requirement and Test debt. Conversely, smells like Organizational Silo Effect (OSE) and Sharing Villainy (SV) show weaker or more isolated correlations. 

\tcbset{colback=blue!7!white,colframe=gray!50!black} 
\begin{tcolorbox}[leftrule=3mm, boxrule=1pt] 
\textbf{Summary for RQ\textsubscript{3}:}
In small projects, OSE, BCE, and TF are strongly linked to Code, Documentation, and Design debt. In medium projects, TF and RS exhibit moderate associations with Defect and Requirement debt. In large projects, TC and PDE emerge as dominant smells, correlating strongly with Code, Design, and Test debt. 
\end{tcolorbox}

\subsection{Answering RQ4: Trend Analysis Between Community Smells and SATD Across ML Projects}

The Mann-Kendall trend analysis reveals evolving trajectories in both community smells and SATD types across ML projects, with notable differences based on project size. While RQ\textsubscript{3} reveals the strength and direction of associations, RQ\textsubscript{4} investigates the temporal dynamics between community smells and SATD across project releases.
For instance, in small projects (P1–P8), we observe a clear presence of statistically significant increasing trends in several community smells such as Organization Silo Effect (OSE), Sharing Villainy (SV), and Radio Silence (RS), particularly in P1–P4. These findings suggest that communication breakdowns, isolated team structures, and uneven knowledge distribution tend to intensify over time, even in relatively small teams. Similarly, almost all SATD types, including Code, Documentation, and Requirement debt, show significant upward trends in most small projects, indicating early-stage accumulating debt.

Medium-sized projects (P9–P16) exhibit mixed trends in community smells and SATD. SV and OSE show strong upward trends, while SV and OS decline in P10, suggesting better collaboration. In P12, most smells show no trend except UI, which decreases significantly. Moreover, four types of SATD, i.e., Design, Documentation, Test, and Requirement debt, showed strong and statistically significant downward trends. However, code debt shows a significant increasing trend, implying that while design- and documentation-related debt is being addressed, implementation-level debt may still be accumulating.

In large-scale projects (P17-P22), Organizational Silo Effect (OSE) frequently trends upward, reflecting persistent structural fragmentation in large teams. The Black Cloud Effect (BCE) and Sharing Villainy (SV) also show a strong rising trend, suggesting growing communication gaps. Conversely, Unhealthy Interaction (UI) and Solution Defiance (SD) often exhibit notable downward trends. However, Despite some improvements in socio-technical factors, all SATD types, except Design debt, continue to rise significantly, suggesting that technical debt persists even as collaboration issues partially improve.

\begin{figure*}
    \centering
    \includegraphics[width=0.8\textwidth]{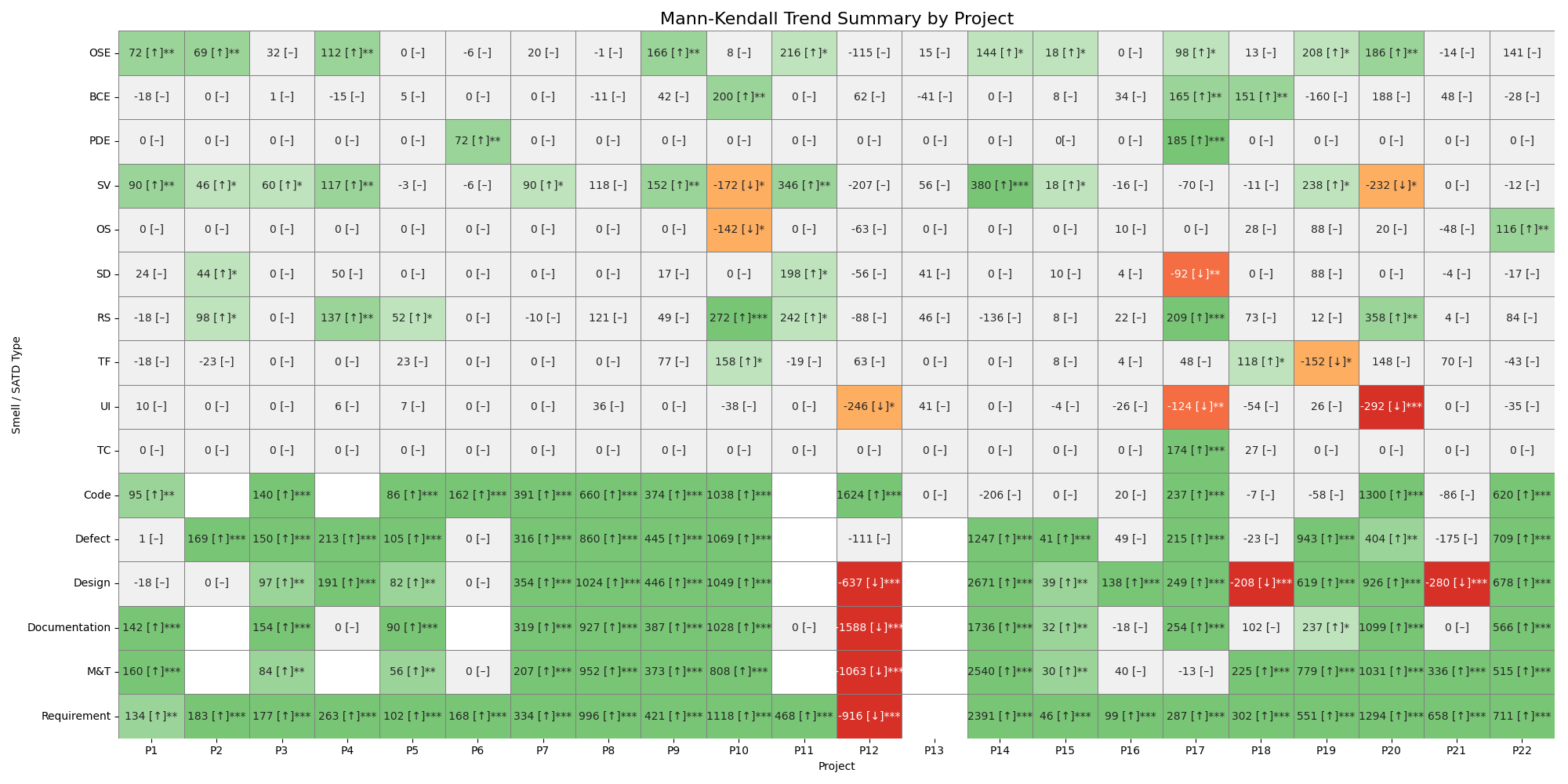}
    \caption{Mann-Kendall trend analysis of community smells and SATD instances over releases }
    \label{fig:enter-label}
\end{figure*}
\tcbset{colback=blue!7!white,colframe=gray!50!black} 
\begin{tcolorbox}[leftrule=3mm, boxrule=1pt] 
\textbf{Summary for RQ\textsubscript{4}:}  Mann-Kendall trend analysis shows that small projects tend to accumulate both community smells (e.g., OSE, SV, RS) and all SATD types over time. Medium projects show mixed trends. However, in large projects, despite reductions in smells like UI and SD, most SATD types (except Design debt) consistently increase. 
\end{tcolorbox}

\section{Key Findings}

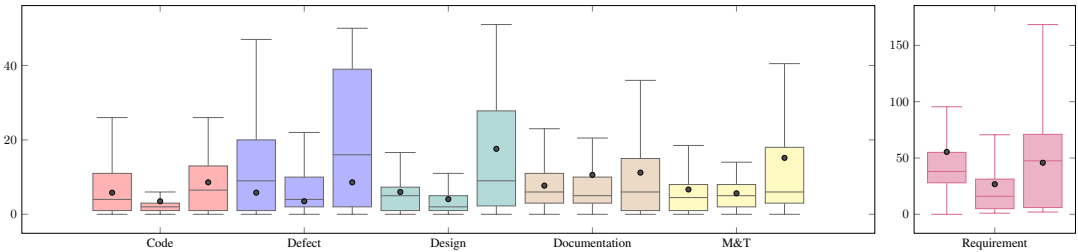
\begin{figure*}[htb]
    \centering
    \resizebox{0.8\textwidth}{!}{
        \subfloat{
            \begin{tikzpicture}
                 \tikzstyle{every node}=[font=\small]
                    \begin{axis}
                        [
                            boxplot/draw direction=y,
                            xtick= {2,5,8,11,14},
                            xticklabels={Code,Defect,Design,Documentation,M\&T},
                            x tick label style={align=center,  font=\small},
                            y tick label style={font=\small},
                            x=1.2cm,
                            custom legend,
                              legend style={
                                legend columns=1,
                                column sep=0.5em,
                                draw=none,
                              }                     
                        ]
                        \addplot+   [ boxplot prepared={lower whisker=0,lower quartile=1.0,median=4,upper quartile=11,upper whisker=26},fill=red!30, draw=black!60] coordinates {};
                        \addplot+   [only marks,mark=*,mark options={scale=0.9},fill=black!70,draw=black!100] coordinates {(1, 5.82)};   
                        \addplot+   [ boxplot prepared={lower whisker=0,lower quartile=1,median=2,upper quartile=3,upper whisker=6},fill=red!30, draw=black!60] coordinates {};
                        \addplot+   [only marks,mark=*,mark options={scale=0.9},fill=black!70,draw=black!100] coordinates {(2, 3.5)}; 
                        \addplot+   [boxplot prepared={lower whisker=0, lower quartile=1.00, median=6.5, upper quartile=13, upper whisker=26},fill=red!30, draw=black!60] coordinates {};
                        \addplot+   [only marks,mark=*,mark options={scale=0.9},fill=black!70,draw=black!100] coordinates {(3, 8.6)};
                        
                        \addplot+   [ boxplot prepared={lower whisker=0,lower quartile=1.0,median=9,upper quartile=20,upper whisker=47},fill=blue!30, draw=black!60] coordinates {};
                        \addplot+   [only marks,mark=*,mark options={scale=0.9},fill=black!70,draw=black!100] coordinates {(4, 5.82)};   
                        \addplot+   [ boxplot prepared={lower whisker=0,lower quartile=2,median=4,upper quartile=10,upper whisker=22},fill=blue!30, draw=black!60] coordinates {};
                        \addplot+   [only marks,mark=*,mark options={scale=0.9},fill=black!70,draw=black!100] coordinates {(5, 3.5)}; 
                        \addplot+   [boxplot prepared={lower whisker=0, lower quartile=2, median=16, upper quartile=39, upper whisker=50},fill=blue!30, draw=black!60] coordinates {};
                        \addplot+   [only marks,mark=*,mark options={scale=0.9},fill=black!70,draw=black!100] coordinates {(6, 8.6)};

                        \addplot+   [ boxplot prepared={lower whisker=0,lower quartile=1.0,median=5,upper quartile=7.3,upper whisker=16.6},fill=teal!30, draw=black!60] coordinates {};
                        \addplot+   [only marks,mark=*,mark options={scale=0.9},fill=black!70,draw=black!100] coordinates {(7, 6)};   
                        \addplot+   [ boxplot prepared={lower whisker=0,lower quartile=1,median=2,upper quartile=5,upper whisker=11},fill=teal!30, draw=black!60] coordinates {};
                        \addplot+   [only marks,mark=*,mark options={scale=0.9},fill=black!70,draw=black!100] coordinates {(8, 4.07)}; 
                        \addplot+   [boxplot prepared={lower whisker=0, lower quartile=2.25, median=9, upper quartile=27.8, upper whisker=51},fill=teal!30, draw=black!60] coordinates {};
                        \addplot+   [only marks,mark=*,mark options={scale=0.9},fill=black!70,draw=black!100] coordinates {(9, 17.6)};

                        \addplot+   [ boxplot prepared={lower whisker=0,lower quartile=3,median=6,upper quartile=11,upper whisker=23},fill=brown!30, draw=black!60] coordinates {};
                        \addplot+   [only marks,mark=*,mark options={scale=0.9},fill=black!70,draw=black!100] coordinates {(10, 7.7)};   
                        \addplot+   [ boxplot prepared={lower whisker=0,lower quartile=3,median=5,upper quartile=10,upper whisker=20.5},fill=brown!30, draw=black!60] coordinates {};
                        \addplot+   [only marks,mark=*,mark options={scale=0.9},fill=black!70,draw=black!100] coordinates {(11, 10.6)}; 
                        \addplot+   [boxplot prepared={lower whisker=0, lower quartile=1, median=6, upper quartile=15, upper whisker=36},fill=brown!30, draw=black!60] coordinates {};
                        \addplot+   [only marks,mark=*,mark options={scale=0.9},fill=black!70,draw=black!100] coordinates {(12, 11.2)};

                        \addplot+   [ boxplot prepared={lower whisker=0,lower quartile=1,median=4.5,upper quartile=8,upper whisker=18.5},fill=yellow!30, draw=black!60] coordinates {};
                        \addplot+   [only marks,mark=*,mark options={scale=0.9},fill=black!70,draw=black!100] coordinates {(13, 6.67)};   
                        \addplot+   [ boxplot prepared={lower whisker=0,lower quartile=2,median=5,upper quartile=8,upper whisker=14},fill=yellow!30, draw=black!60] coordinates {};
                        \addplot+   [only marks,mark=*,mark options={scale=0.9},fill=black!70,draw=black!100] coordinates {(14, 5.64)}; 
                        \addplot+   [boxplot prepared={lower whisker=0, lower quartile=3, median=6, upper quartile=18, upper whisker=40.5},fill=yellow!30, draw=black!60] coordinates {};
                        \addplot+   [only marks,mark=*,mark options={scale=0.9},fill=black!70,draw=black!100] coordinates {(15, 15.17)};

                    \end{axis}
                \end{tikzpicture}
            }
        \subfloat{
            \begin{tikzpicture}
                 \tikzstyle{every node}=[font=\small]
                    \begin{axis}
                        [
                            boxplot/draw direction=y,
                            xtick= {2},
                            xticklabels={Requirement},
                            x tick label style={align=center,  font=\small},
                            y tick label style={font=\small},
                            x=1.2cm
                        ]
                        \addplot+   [ boxplot prepared={lower whisker=0,lower quartile=28,median=38,upper quartile=55,upper whisker=95.5},fill=purple!30, draw=purple!60] coordinates {};
                        \addplot+   [only marks,mark=*,mark options={scale=0.9},fill=black!70,draw=black!100] coordinates {(1, 55.42)};   
                        \addplot+   [ boxplot prepared={lower whisker=1,lower quartile=5,median=16,upper quartile=31.25,upper whisker=70.63},fill=purple!30, draw=purple!60] coordinates {};
                        \addplot+   [only marks,mark=*,mark options={scale=0.9},fill=black!70,draw=black!100] coordinates {(2, 26.76)}; 
                        \addplot+   [boxplot prepared={lower whisker=2, lower quartile=6, median=47.5, upper quartile=71, upper whisker=168.5},fill=purple!30, draw=purple!60] coordinates {};
                        \addplot+   [only marks,mark=*,mark options={scale=0.9},fill=black!70,draw=black!100] coordinates {(3, 45.8)};
                    \end{axis}
                \end{tikzpicture}
            }
        }
    
    \caption{Quartile distribution of SATD types by project size (small, medium, large)}
    \label{fig:satd-growth}
\end{figure*}

\textbf{Scaling Project, Swelling Debt.} Prior work by Tamburri et al. \cite{tamburri2019exploring} showed that community smells like OSE, PDE, and BCE are prevalent in open-source communities. Our findings from \textbf{RQ\textsubscript{1}} extend this to ML projects, where BCE, PDE, RS, TF, and TC consistently appear. In contrast, OSE and SV emerge mainly in larger teams, suggesting that subgroup formation and socio-technical debt increase with project size. This pattern aligns with Almarimi et al. \cite{Almarimi2020LearningTD}, who linked turnover and communication gaps to community smell emergence.
Further analysis of SATD distributions in Fig.~\ref{fig:satd-growth} across project sizes verifies this observation. Our quartile analysis reveals that as ML projects grow in size, the presence and severity of SATD also increase. While small projects tend to accumulate a disproportionately high amount of requirement debt, medium and large projects experience sharp increases in defect, design, documentation, and test-related debt. Notably, large projects show high variability and upper quartiles across nearly all SATD types, suggesting that socio-technical debt manifests more broadly and intensively as teams scale. This finding strengthens the argument that the emergence of community smells is not only a symptom of poor social structure but also a potential precursor to SATD accumulation, particularly in complex ML ecosystems. Consequently, ML project leads should address community smell indicators and implement early and size-aware strategies to monitor and mitigate SATD growth over time.

\textbf{Command, Conflict and Code Decay.} 
\begin{table*}[]
    \centering
    \caption{Examples of community smells and SATD co-existence}
    \normalsize
    \resizebox{0.9\textwidth}{!}{
        \begin{tabular}{l|p{4cm}|p{18cm}}
        \toprule
        Project     &   SATD Example   &    Observation  \\ 
        \midrule
        P8          & \textcolor{magenta}{\textit{TODO PK do this once during construction so that we don't have to do it again}}, Design Debt
                    &   This project had the SATD instance persisted from release 1.20 to 2.7, a period during BCE was constantly detected. Notably, BCE was no longer present in 2.7 and following this, the SATD was removed from the project. This temporal pattern suggest that resolving BCE-related socio-technical challenges, such as improved team communication and knowledge sharing enabled delayed design improvements.
                    \\ \midrule
        P10         &   \textcolor{magenta}{\textit{todo why does it only work when we render a mesh before\>?}}, Defect Debt
                    &   This SATD instance persisted from release 1.1 to 2.2, during which the Truck Factor (TF) community smell was consistently detected. Following release 2.2, both the SATD and TF smell disappeared simultaneously. This pattern suggests that high TF can delay technical debt resolution, and that distributing team knowledge helps address lingering SATD, reinforcing the link between socio-technical bottlenecks and debt persistence. \\ \midrule
        P19         &   \textcolor{magenta}{\textit{TODO: xv and yv not used, should the next line be changed to xv, yv?}}, Code Debt
                    & We observed that this SATD instance persisted across releases 19.1 to 21.0, during which the Radio Silence (RS) community smell was also consistently detected. However, starting from release 22.0 and continuing through 23.0, RS was no longer present, and this SATD instance was also removed. This temporal alignment suggests that the resolution of RS potentially through improved communication and collaboration practices may have facilitated the identification and removal of that SATD. \\
        \bottomrule
        \end{tabular}
    }
    \label{tab:coexistence}
    \vspace{-1em}
\end{table*}
Our findings from \textbf{RQ\textsubscript{2}} and \textbf{RQ\textsubscript{3}} empirically substantiate the theoretical claim articulated by Serebrenik \cite{serebrenik2022social}, who hypothesized that specific community smells foster the introduction of SATD or reduce the likelihood of its elimination. Our project level analysis in \textbf{RQ\textsubscript{2}} shows that Radio Silence, Sharing Villainy and Organizational Silo Effect are the most reliable predictors of higher SATD counts across the ML projects which indicates their role as debt ``originator". Furthermore, our results from \textbf{RQ\textsubscript{3}} reveal that in larger projects, authority-centered smells like the Prima-donna Effect and interaction-based smells such as Toxic Communication are significantly correlated with persistent code, design, and maintenance debt. This correlation can be attributed to the socio-technical dynamics prevalent in various development teams. In case of PDE, dominant individuals may centralize decision-making, resist collaborative input, and become bottlenecks for knowledge dissemination \cite{tamburri2015social}. Such behavior can lead to architectural decisions that are not broadly evaluated, resulting in design choices that are difficult to modify or refactor. Similarly, TC creates a hostile communication environment, discouraging team members from engaging in open discussions or raising concerns about code quality \cite{sayago2025analyzing}. This atmosphere can suppress proactive maintenance activities and the early identification of issues, allowing SATD to persist and grow. The findings from our study underscore the need for inclusive decision-making and communication platforms. Moreover, project managers should foster structured code reviews and establish psychological safety to encourage open dialogue, mitigating TC’s impact on the occurrence of SATD. 

Additionally, we also attempted to find some examples of the co-existence of community smells and SATD from our studied projects. Table~\ref{tab:coexistence} illustrates some scenarios and our observations, which signify the potential interplay between socio-technical issues and the persistence of technical debt. These examples motivate the need for practitioner-centered validation to deepen the understanding of these dynamics.

\begin{figure}
    \centering
    \includegraphics[width=0.85\linewidth]{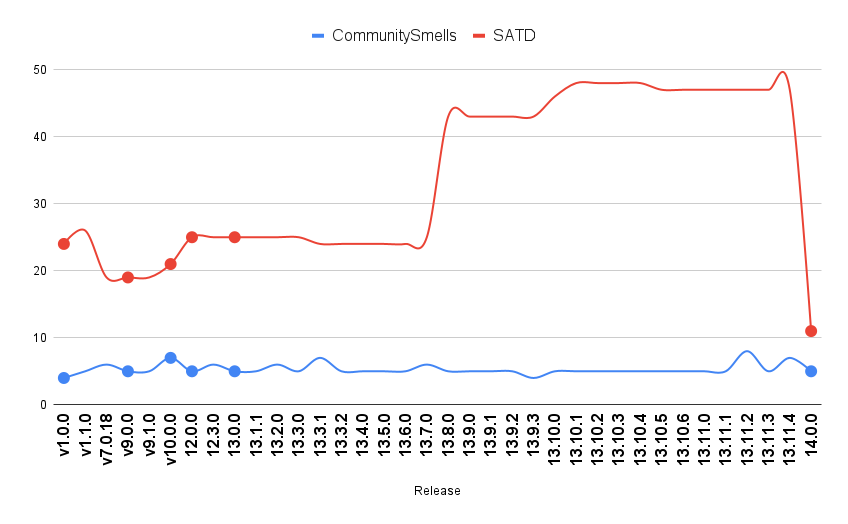}
    \caption{Trend of community smells and SATD over major and minor releases in \textit{P12}}
    \vspace{-1em}
    \label{fig:releases}
    \vspace{-1em}
\end{figure}

\textbf{Spikes Before the Storm, Reset at the Release.} Our findings from \textbf{RQ\textsubscript{4}} reveal the temporal evolution of community smells and SATD instances across project releases. Fig.~\ref{fig:releases} illustrates a clear similarity in the trends of community smells and SATD of the project \textit{P12}, particularly around major releases. To better understand these trends, we analyzed the data, distinguishing between major and minor releases, classifying releases as major when adhering to the semantic versioning pattern \textit{x.0.0} \cite{wu2023characterize}.
In project P12, we observed a notable combined upward and downward trend. Initially, both community smells and SATD showed relatively moderate and stable levels. However, around release 13.8.0, a significant increase occurred, particularly evident in SATD, which rose sharply and remained persistently high across subsequent minor releases. During this elevated SATD phase, there was also a noticeable increase in community smells, indicating concurrent socio-technical challenges related to collaboration and structural organization.
Importantly, release 14.0.0 marked a clear inflection point, with both SATD and community smells sharply declining. This simultaneous reduction suggests that deliberate interventions, likely involving extensive refactoring, improved documentation, or structural adjustments, were undertaken. The temporal rise and fall of both metrics highlight their interconnection. Our findings from \textbf{RQ\textsubscript{4}} suggest that intensive development phases often fuel parallel growth in technical and social debt, while major releases offer key opportunities to reduce both and enhance project quality.

\section{Threats to Validity}
Threats to \textbf{construct validity} involve the alignment between theory and observation. Our study relies on the detection capabilities of automated tools, csDetector for community smells, and MT-BERT-SATD for identifying SATD. While both tools have demonstrated strong performance in prior work, they are inherently limited by their trained heuristics and model architectures. csDetector may miss smells occurring in off-platform or informal communication, and the SATD classifier may mislabel comments that do not follow conventional SATD phrasing. To address this, we conducted robustness checks and manual validation on a subset of projects to ensure the reliability of our findings (Section~\ref{methodology-csDetector}). Nonetheless, we acknowledge that not all community smells or SATD instances may be captured, and our results reflect patterns detectable by the tools. To maintain clarity, we explicitly refer to our findings as ``community smells detectable by csDetector,” recognizing the tool's scope and limitations.

Threats to \textbf{internal validity} involve factors within the study that may affect results. We aligned community smells and SATD using release-based analysis, assuming they co-evolve within the same cycle. However, delays in their interaction may exist. The csDetector's ML-based smell detection may introduce slight variability, which we mitigated through multiple runs and manual validation on a subset of projects.

Threats to \textbf{external validity} concern the generalizability of our findings. Since our dataset is limited to Github-hosted projects, it may not capture collaboration patterns common in corporate environments. Moreover, while our dataset broadly targets ML systems, it does not isolate deep learning-specific projects. Prior work by Pepe et al. \cite{pepe2024taxonomy} suggests that DL systems often introduce unique forms of SATD such as fragile data pipelines or model retraining constraints, that may not be fully addressed by general-purpose classifiers like MT-BERT-SATD. However, our use of 155 active, high quality ML projects curated by Latendresse et al. \cite{latendresse2024exploratory} strengthens the reliability of our results to a great extent and offers a solid empirical basis for our findings.

Threats to \textbf{conclusion validity} concern the soundness of statistical and logical inferences. We used Spearman’s rank and Point-Biserial correlation to examine links between community smells and SATD. While suitable for our data, these methods reveal co-occurrence, not causation. Temporal trends suggest alignment but not directionality. Future work should explore causal models to validate these patterns.

\section{Conclusion and Future Work}
In this study, we conducted an empirical investigation into the relationship between community smells and SATD in 155 open-source ML-based software projects. Our analysis revealed that community smells are not only prevalent in ML systems but also closely associated with the accumulation and persistence of SATD. We identified specific smells, such as Radio Silence, Organizational Silo, and Sharing Villainy, as strong indicators of technical debt, particularly in requirements, design, and documentation. Furthermore, our release-level trend analysis shows that both community smells and SATD tend to escalate as projects scale, with large projects exhibiting high variability and persistence across most debt categories. These findings highlight the importance of tracking socio-technical indicators over time and emphasize the value of targeted interventions at major release points.

Future work includes developer interviews to uncover contextual factors behind community smells, and investigating how resolving these smells affects SATD reduction. We also plan to improve detection using advanced LLMs and expand our analysis to industrial and non-GitHub ML projects for broader insights into socio-technical debt.

\section{Acknowledgement}
This research is supported in part by the Natural Sciences and Engineering Research Council of Canada (NSERC) Discovery Grants program, the Canada Foundation for Innovation's John R. Evans Leaders Fund (CFI-JELF), and by the industry-stream NSERC CREATE in Software Analytics Research (SOAR).

\bibliographystyle{IEEEtran}
\bibliography{ref}

\end{document}